\documentclass[12pt]{amsart}
\usepackage{graphicx}
\usepackage{amssymb}

\newcommand{\R}{\mathbb R}
\newcommand{\C}{\mathbb C}
\newcommand{\support}{{\rm supp\,}}

\renewcommand{\Re}{ {\rm Re}}
\renewcommand{\Im}{{\rm Im}}

\newtheorem{theorem}{Theorem}
\newtheorem{corollary}[theorem]{Corollary}
\newtheorem{proposition}[theorem]{Proposition}

\newtheorem{lemma}[theorem]{Lemma}

\newcommand{\thetac}{\pi/3}

 \title[Logarithmic equilibrium measure]{The support of the logarithmic equilibrium measure on sets of revolution
 in $\R^3$}
 \author{D. P. Hardin$^{1}$, E. B. Saff$^{2}$,  and H. Stahl$^{3}$} 
\address{D. P. Hardin and E. B. Saff: Center for Constructive Approximation, Department of Mathematics,  
Vanderbilt University,
Nashville, TN 37240, USA }
\email{Doug.Hardin@Vanderbilt.Edu}
\email{Edward.B.Saff@Vanderbilt.Edu}
\address{H. Stahl: TFH-Berlin/FBII, Luxemburger Strasse 10 13353 Berlin, Germany }
\email{stahl@tfh-berlin.de}
\begin{document}

\thanks{ \hspace*{-.26in} $^{1}$The research of this author  was supported, in part,
by the U. S. National Science Foundation under grants DMS-0505756 and DMS-0532154. \\
$^{2}$The research of this author  was supported, in part,
by the U. S. National Science Foundation under grant DMS-0532154.\\
$^{3}$The research of this author  was supported, in part,
by INTAS Research Network NcCCA 03-51-6637.}

\keywords{Potential, Equilibrium measure, Logarithmic potential, Surfaces of revolution,
Riesz energy}
\subjclass{Primary 11K41, 70F10, 28A78; Secondary 78A30, 52A40}

\begin{abstract}
For surfaces of revolution $B$ in $\R^3$, we investigate the limit
distribution of minimum energy point masses on $B$ that interact according
to the logarithmic potential $\log (1/r)$, where $r$ is the Euclidean distance 
between points.  We show that such limit distributions are supported only on
the ``out-most'' portion of the surface (e.g., for a torus, only on that portion of the 
surface with positive curvature). Our analysis proceeds by reducing the problem to the
complex plane where a non-singular potential kernel arises whose level lines 
are ellipses. 
\end{abstract}

\maketitle

\section{Introduction}

For a collection of $N(\ge 2)$ distinct points
$\omega_N : =\{x_1,\ldots, x_N\}\subset \R^3$ and $s> 0$,
 the  {\em   Riesz $s$-energy of $\omega_N$} is defined by
$$
E _s(\omega_N) := 	\sum_{1\leq i\neq j\leq N}{k_s(x_i,x_j)} =\sum_{i=1}^N\sum_{ {j=1} \atop { j\neq i}}^Nk_s(x_i,x_j),
 $$
 where, for $x,y \in \R^3$,  
$
k _s(x,y):=  	 
{1}/{\left |x-y\right|^s} $.
As $s\to 0$, it is easily verified that $$(k_s(x,y) -1)/s \to \log\left({1}/{\left|x-y\right|}\right)$$ and so it is
natural to define $ k_0(x,y):= \log\left( {1}/{\left| x-y\right|}\right)$. 
 For a compact set  $B\subset \R^3$ and $s\ge 0$,
  the {\em $N$-point
   $s$-energy of $B$}  is defined by
\begin{equation} \label{Es}\mathcal E_s(B,N):=\inf \{E_s(\omega_N) \mid  \omega_N\subset B ,
 |\omega_N |=N\},
\end{equation} where $|X|$ denotes
the cardinality of a set $X$.   
Note that the logarithmic ($s=0$) minimum energy  problem is equivalent
to the  maximization of the product
$$
\prod_{1\leq i\neq j\leq N}{\left|x_i-x_j\right|},
$$
and  that for planar sets, such optimal points are known as  {\it Fekete points}.    
(The fast generation of near optimal logarithmic energy points for the sphere $S^2$
is the focus of one of S. Smale's   ``mathematical problems for the next century'';
see \cite{S}.)

 \begin{figure}[tbp]
\begin{center}
\hspace*{-.5in}
\includegraphics[width=6in]{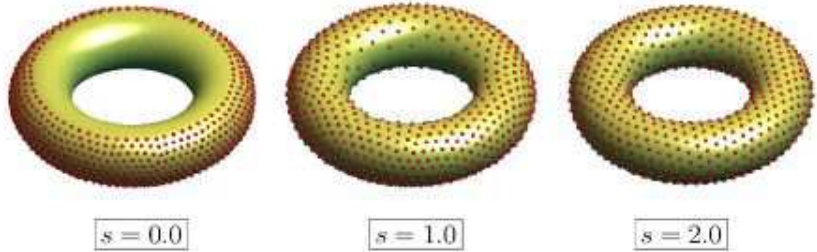}
\vspace*{-5.8in}
\caption{Near optimal Riesz $s$-energy configurations ($N=1000$ points) on a torus in $\R^3$ for $s=0,1,$ and 2.}
\label{fig1}
\end{center}
\end{figure}

If $0\le s< {\rm dim}\ B$ (the Hausdorff dimension of  $B$),  the  limit distribution (as $N\to \infty$) of   optimal
$N$-point configurations is given by the {\em equilibrium measure} $\lambda_{s,B}$
that minimizes the continuous
energy integral
$$
I_s(\mu):= \iint_{B\times B}  k_s (x,y) \ d\mu (x)\ \! d\mu (y) 
$$
over the class $\mathcal{M}(B)$ of (Radon) probability measures $\mu$ supported on $B$.  In addition,  the asymptotic order of the Riesz $s$-energy is $N^2$;   more precisely we have  $\mathcal E_s(B,N)/N^2\to I_s(\lambda_{s,B})$ as $N\to \infty$   (cf. \cite[Section II.3.12]{L}).
 In the case when $B=S^2$, the unit sphere in $\R^{3}$, the equilibrium measure is simply the normalized surface area measure.
  If $s\geq {\rm dim }\ B$, then $I_s(\mu)=\infty$ for every $\mu\in \mathcal{M}(B)$ and potential
  theoretic methods cannot be used. However, it was recently
  shown in    \cite{HarSaf04} that
  when $B$ is a  $d$-rectifiable manifold  of positive $d$-dimensional Hausdorff measure and $s\ge d$,
   optimal $N$-point configurations are uniformly distributed (as $N\to\infty$) on $B$ with respect to $d$-dimensional Hausdorff measure restricted to $B$.  (The assertion for the case $s=d$ further requires that $B$ be a subset of a $C^1$ manifold.)      For further extensions of these results, see   \cite{BorHarSaf05}.  Related results 
and applications appear in  \cite{ConSlo99} (coding theory),
  \cite{SloWom} (cubature on the sphere), and  
 \cite{BF} (finite normalized  tight frames).

\begin{figure}[tbp]
\begin{center}
 \includegraphics[width=\textwidth]{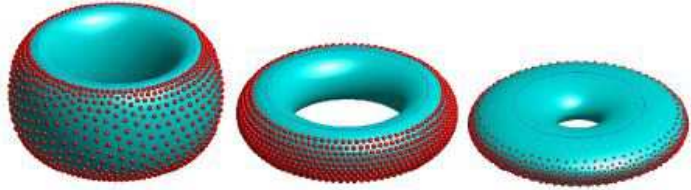}
 
 \vspace*{-4.5in}
\caption{Minimum logarithmic energy points on various toroidal surfaces.  }
\label{fig2}
\end{center}
\end{figure}

In Figure~\ref{fig1}, we show near optimal  Riesz $s$-energy configurations for the values of $s=0,1,$ and $2$ for $N=1000$ points
restricted to live on the torus $B$ obtained by revolving the circle of radius 1 and center
$(3,0)$ about the $y$-axis.  (For recent results on the disclinations of minimal energy 
points on toroidal surfaces, see \cite{BNT04}.)   The somewhat surprising observation that there are no points on the ``inner'' part of the
torus in the case $s=0$ (and, in fact, as well for $s$ near 0) is what motivated us to investigate the support of the logarithmic equilibrium measure $\lambda_{0,B}$.     In this paper we show that, in fact, this is a general 
phenomenon for optimal logarithmic energy configurations of points restricted to sets  of revolution in $\R^3$ (see Figure~\ref{fig2}).

\section{Preliminaries}

 In this paper we focus on the logarithmic
kernel $k_0$.  Let  $B\subset \R^3$ be compact.  
As in the previous section, the {\em logarithmic energy} of a measure $\mu\in \mathcal{M}(B)$  is given by
\begin{equation}\label{3dkernel}
 I_0(  \mu)=\iint_{B\times B} \log \frac{1}{|p-q|}\, d \mu(p)
d \mu(q) 
\end{equation}
and the corresponding {\em potential $U^{\mu} $} is defined by 
\begin{equation}\label{3dpotential}
U^{\mu} (p):=\int_{B} \log \frac{1}{|p-q|}\, d \mu(q) \qquad (p\in \R^3). 
\end{equation}
Let $V_B:=\inf_{ \mu\in \mathcal{M}(B)}I_0( \mu)$.  
The {\em logarithmic capacity of $B$}, denoted by ${\rm cap } (B)$, is exp$(-V_B)$.  A condition $C(p)$ is said to hold 
{\em quasi-everywhere} on $B$ if it holds for all $p\in B$ except for a subset of logarithmic capacity zero.\footnote{
The logarithmic capacity of a Borel set $E$ is the sup of the capacities of its compact subsets.
Any set that is contained in a Borel set of capacity zero is said to have capacity zero.} 
If ${\rm cap  }(B)>0$, then  there is a unique probability measure $ \mu_{B}\in \mathcal{M}(B)$ 
(called the {\em equilibrium measure on $B$}) such that $ I(  \mu_{B})=V_B$ (this is implicit in the 
references  \cite{L,ST}).     
Furthermore, the equality $U^{  \mu_{B}}(p)=  V_{B}$ holds quasi-everywhere on the  support of $\mu_{B}$ and $U^{\mu_B}(p)\ge   V_{B}$ quasi-everywhere on $B$.

We now turn our attention to sets of revolution in $\R^3$.  
Let $\R_+:=[0,\infty)$ and, for $t\in [0,2\pi)$, let   $\sigma_t:\R^3\to\R^3$ denote the rotation about the $y$-axis through an angle $t$: $$\sigma_t(x,y, \zeta)=(x\cos t-\zeta \sin t,y,x\sin t+\zeta \cos t). $$ 
 For a   compact set  $A$ contained in the right half-plane $ H^+:=\R_+\times \R$,  let 
 $\Gamma (A)\subset \R^3$
 be the
set obtained by revolving $A$ around the $y$-axis, that is, 
$$\Gamma(A):=\{\sigma_t(x,y,0)\, \mid  \, (x,y)\in A,\, 0\le t <2 \pi\}.$$
We say that  $A\subset H^+$ is {\em non-degenerate }
if ${\rm cap \,}(\Gamma(A))$ is positive.  For example,
if $A$ contains at least one point not on the $y$-axis, then $A$ is non-degenerate.

  \noindent

\section{Reduction to the $xy$-plane}
   A Borel measure $\tilde \nu \in \mathcal{M} (\R^3)$ is {\em rotationally symmetric about the $y$-axis}
if  $\tilde\nu=\tilde \nu \circ \sigma_t  $ for all  $ t\in [0,2\pi)$.  If $\tilde \nu$ is rotationally symmetric about the $y$-axis, then
  $d \tilde \nu =\frac{1}{2\pi}dt  d\nu$,
 where  $\nu: = \tilde \nu\circ \Gamma\in \mathcal{M}(H^+)$ and $dt$ denotes Lebesgue measure
 on $[0,2\pi)$. 
 Identifying points  $z,w \in H^+$ as complex numbers    
 $z=x+iy=(x,y,0)$ and $w=u+iv=(u,v,0)$ we have  
  \begin{eqnarray} \label{IJ}
 I_0(\tilde \nu)&=&\iint_{\R^3\times \R^3}\log \frac{1}{|p-q|}\, d\tilde \nu(p)d \tilde \nu(q) \nonumber \\
&=&  \iint_{H^+\times H^+}K(z,w) \,   d \nu(z)   d\nu(w)  \\
&=:&J(\nu),\nonumber
\end{eqnarray}
where 
\begin{equation}
\label{Kdef}
K(z,w):= \frac{1}{2\pi} \int_{0}^{2 \pi}  \log \frac{1}{|\sigma_t(z)-w|}\;   dt.
\end{equation}
Notice that
\begin{eqnarray}\label{ker1}
|\sigma_t(z)-w|^2&=&(x\cos t-u)^2 +(y-v)^2+x^2\sin^2 t \\ &=& x^2+u^2+(y-v)^2-2xu \cos t. \nonumber
\end{eqnarray}
Let  $w_*:=-u+iv=-\overline w$ denote the reflection of $w$ in the $y$-axis.
Then, using (\ref{ker1}) and  the formula
$$
\frac{1}{2\pi}\int_0^{2\pi} \log(a+b \cos t)\ dt=\log \frac{a+\sqrt{a^2-b^2}}{2} 
$$
with  
$a=(y-v)^2+x^2+u^2$ and $b=-2xu$,   we obtain 
\begin{equation}
\label{kernel}                       
K(z,w) =-\frac{1}{2}\log \frac{a+\sqrt{a^2-b^2}}{2} 
=\log \frac{2}{|z-w|+|z-w_*|} , 
\end{equation}
 where we have used  $$2\left(a+\sqrt{a^2-b^2}\right)=\left(\sqrt{a+b}+\sqrt{a-b}\right)^2=\left(|z-w|+|z-w_*|\right)^2.$$  
 
 \subsection{Equilibrium measure   $\lambda_A\in \mathcal{M}(A)$}
For a non-degenerate compact set $A\subset H^+$,
 the uniqueness of the equilibrium measure $\mu_{\Gamma(A)}$ 
and the symmetry of the revolved set $\Gamma(A)$  imply that  $\mu_{\Gamma(A)}$
is rotationally symmetric about the $y$-axis 
    and so  $d \mu_{\Gamma(A)}= \frac{1}{2\pi}dt  d\lambda_A$,
 where for any Borel set $B\subset H^+$ 
 \begin{equation}\label{lamAdef}
 \lambda_A(B) :=  \mu_{\Gamma(A)}( \Gamma(B)).
 \end{equation}
 Furthermore, if $\nu\in \mathcal{M}(A)$, then $d \tilde \nu: =\frac{1}{2\pi}dt  d\nu$
 is rotationally symmetric about the $y$-axis and so we  have
 $$J(\lambda_A)\ge \inf_{\nu \in \mathcal{M}(A)}J(\nu) =\inf_{\nu \in\mathcal{M}(A)}I_0(\tilde \nu)
\ge  I_0(\mu_{\Gamma(A)})=J(\lambda_A), $$   
which leads to the following proposition.
 \begin{proposition}\label{prop1}
Suppose $A$ is a  non-degenerate compact set in $H^+$ and let $\lambda_A\in \mathcal{M}(A)$ be 
defined by (\ref{lamAdef}).  Then $\lambda_A$ is the unique measure in $\mathcal{M}(A)$
that minimizes $J(\nu)$ over all measures $\nu\in \mathcal{M}(A)$.   
That is, $\lambda_A$ is the {\em equilibrium measure} for the kernel $K$ and set $A$.
\end{proposition}


For $\nu\in \mathcal{M}(A)$, we define the {\em ($K$-)potential $W^\nu$ } by
\begin{eqnarray}\label{Kpot}
W^\nu(z)&:=&\int_AK(z,w)\, d\nu(w)\\
&=&\int_A \log \frac{2}{|z-w|+|z-w_*|} d\nu(w)\qquad (z\in H^+).\nonumber
\end{eqnarray}
Then, for $z=(x,y,0)\in H^+$,  we have
\begin{eqnarray*}
U^{\mu_{\Gamma(A)}}(z)&=&\int_{\Gamma(A)}\log\frac{1}{|z-q|}\, d\mu_{\Gamma(A)}(q)\\
&=&\frac{1}{2\pi}\int_{A} \int_{0}^{ 2\pi}  \log  \frac{1}{|z-\sigma_t(w)|} \,  dt\, d\lambda_A(w)\\
&=&\int_AK(z,w)\, d\lambda_A(w) =W^{\lambda_A}(z).
\end{eqnarray*}
From the properties of $U^{\mu_{\Gamma(A)}}$, we then infer the following lemma.

\begin{lemma}\label{lampotconstant}
Suppose $A $ is a  non-empty compact set in the interior of $H^+$.  Let $\lambda_A$ be the equilibrium measure for $A$ with respect to the kernel $K$. Then
 the potential $W^{\lambda_A}$ satisfies 
 $W^{\lambda_A}(z)=J(\lambda_A)$  
 for $z$ in the support of $\lambda_A$ and $W^{\lambda_A}(z)\ge J(\lambda_A)$ for $z\in A$. 
\end{lemma}
\noindent
{\bf Remark:}
In Lemma~\ref{lampotconstant} we no longer need a quasi-everywhere exceptional set, since each
point of $A$ generates a circle in $\R^3$ with positive logarithmic capacity. 

\subsection{Properties of $K$.}

\begin{figure}[tbh]
\vspace{.2in}
\centerline {
\includegraphics[width=3in]{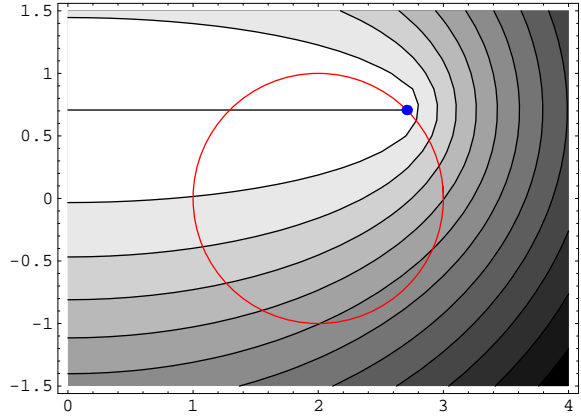}
}
\vspace{.2in}
\caption{\label{levelcurves} Level curves for $K(z,w)$ for $w$ a fixed point on the unit circle centered at (2,0).}
\end{figure}

Let $s(z,w):=|z-w|+|z-w_*|$.  Then $K(z,w)=-\log (s(z,w)/2)$ 
and so, for fixed $w\in H^+$, the level sets  of  $K(\cdot,w)$  are ellipses with foci  $w$ and $w_*$ as shown in Figure~\ref{levelcurves}.  Since the foci have the same imaginary part $v=\Im[w]=\Im[w_*]$, it  follows from geometrical considerations that $K(\cdot,w)$ is 
strictly decreasing along horizontal rays $[iy,\infty+iy)$ for $y\neq v$.  Along the horizontal ray $[iv,\infty+iv)$, we
have that $K(\cdot,w)$ is constant on  the line  segment $[iv,w]$ and  
strictly decreasing on the   ray $[w,\infty+iv)$.    

Furthermore, $K$ is clearly continuous at any $(z,w)\in H^+\times H^+$ unless
$z=w=iy$ for some $y\in\R$.  Since $|z-w_*|=|(z-w_*)_*|=|w-z_*|$, it follows that $K$ is symmetric, that is, $K(z,w)=K(w,z)$ for $z,w\in H^+$.  We summarize these properties of $K$ in the following lemma.

\begin{lemma}\label{Kprop}
The kernel $K:H^+\times H^+\to \R $   in  (\ref{kernel}) has the following properties:
\begin{enumerate}
\item[(a)]  $K$ is symmetric:  $K(z,w)=K(w,z)$ for $w,z\in H^+$.
\item[(b)] $K$ is continuous at all points $(z,w)\in H^+\times H^+$
except  points $(z,z)$ such that $\Re(z)=0$. 
\item[(c)]  Let $u\ge 0$ and $y\neq v\in \R$ be fixed. Then  $K(x+iy,u+iv)$ is a strictly decreasing function  of $x$ for $x\in [0,\infty)$. Furthermore, 
  $K(x+iy,u+iy)$ is constant  
for $x\in [0,u]$ and is strictly decreasing for $x\in [u,\infty)$.  
\end{enumerate}
\end{lemma}



The following lemma is then a consequence of  Lemma~ \ref{Kprop}. \begin{lemma}\label{potdec}
Suppose $\nu\in \mathcal{M}(A)$ is not a point mass (that is, the support of $\nu$ contains at least two points).  Then the potential
$W^\nu(z)$ is strictly decreasing along the horizontal rays $[iy,\infty+iy)$
for all $y\in \R$.  
\end{lemma}

If $A$ is a  non-degenerate compact set in $H^+$, let  $P (A)$ denote the projection of the set $A$ onto the $y$-axis and for $y\in P(A)$, define  $x_A(y)=\max  \{ x \, \mid  \, (x,y)\in A\}$.  We then let  $A_+$ denote the ``right-most''
portion of $A$, that is,  $$A_+:=\{(x_A(y),y) \mid  y\in P(A)\}.$$

Using Lemmas~\ref{lampotconstant} and \ref{potdec} we then obtain the following result.

\begin{theorem}\label{Aplus}
Suppose $A$ is a   compact set in $H^+$   such that $A_+$ is contained in the interior of $H^+$. Then 
the support of the  equilibrium measure $\lambda_A\in \mathcal{M}(A)$ is contained in $ A_+$.  
\end{theorem}


%

\section{Convexity}
Recall that a function $f:[a,b]\to \R$ is {\em strictly convex  on $[a,b]$} if
$f(\theta x+(1-\theta )y)< \theta f(x)+(1-\theta)f(y)$ for all $a\le x<y\le b$ and
$0<\theta<1$.  
\begin{theorem} \label{convexlemma1}
Suppose $A$ is  a    compact set in $H^+$  such that
$A_+$ is contained in the interior of $H^+$ and $\gamma:[a,b]\to H^+$ is   continuous.
Further suppose that  
  \begin{enumerate}
  \item[(a)] $A_+\subset \gamma^*:=\{\gamma(s) \mid a\le s\le b\}$ and
\item[(b)] $K(\gamma(\cdot),\gamma(s))$ is a strictly convex function on the intervals
$[a,s]$ and $[s,b]$ for each fixed $s\in [a,b]$.
\end{enumerate}
Then there is some closed interval $I\subset [a,b]$ such that
$ {\rm supp\,  }\lambda_A = \gamma(I)\cap A_+$.  
 \end{theorem}
 \begin{proof}
Suppose $A$ and $\gamma$ satisfy (a) and (b).  From  Theorem~\ref{Aplus}
we have $ \support\lambda_A\subset \gamma^*$.  
Let $t_1:=\min_{a\le t\le b} \{t \mid \gamma(t )\in   \support\lambda_A\}$
and 
$t_2:=\max_{a\le t\le b} \{t \mid \gamma(t )\in   \support\lambda_A\}$.
Suppose that $G$ is an open interval in $I:=[t_1,t_2]$ such that
$\gamma(G)\cap \,  \support\lambda_A=\emptyset$.  
Then $W^{\lambda_A}\circ \gamma$ is strictly convex on $G$  and 
$W^{\lambda_A}(z)=J(\lambda_A)$ for $z\in    \support\lambda_A$ and so we have
$W^{\lambda_A}(\gamma(t))<J(\lambda_A)$
 for $t\in G$.  Hence,  Lemma~\ref{lampotconstant} implies that
 $\gamma(G)\cap A =\emptyset$ which then implies $ \support\lambda_A = \gamma(I)\cap A_+$.
\end{proof}

We next consider several examples where we can verify that the hypotheses of
Theorem~\ref{convexlemma1}   hold.   In these examples, $\gamma$ is a smooth curve,
 but note that $A_+$ is only required to be a compact subset of $\gamma^*$.  For
 example, $A_+$ may be a Cantor subset of $\gamma^*$.
 
We first consider a  case where we can completely specify the support of
$\lambda_A$.  

\begin{corollary}\label{vertprop}
Suppose $A$ is a  non-degenerate compact subset in  $H^+$   such that $A_+$ is contained in a
 vertical line segment $[R+ci,R+di]$ for some $R>0$.  
 Then $ \support\lambda_A =  A_+$.
 \end{corollary}
 \begin{proof}
  Consider the parametrization 
 $\gamma(t)=R+i t$, $c\le t\le d$, of the line segment $[R+ci,R+di]$.  For  $s,t\in [c,d]$, $s\neq t$, direct calculation shows 
 $K(\gamma(t),\gamma(s))=-\log(|s-t|+\sqrt{4 R^2+(s-t)^2})+\log 2$ and 
  \begin{eqnarray} \label{vert1}
 \frac{d}{dt} K(\gamma(t),\gamma(s))&=&\frac{\text{sgn}(s-t)}{\sqrt{4 R^2+(s-t)^2}}, \\
 \frac{d^2}{dt^2} K(\gamma(t),\gamma(s))&=& \frac{|s-t|}{\left(4 R^2+(s-t)^2\right)^{3/2}}.  \label{vert2}
 \end{eqnarray}
  Then (\ref{vert2}) shows that condition (b) of Theorem~\ref{convexlemma1} holds and therefore  there is some
 interval $I=[t_1,t_2]$ such that $ \support\lambda_A = \gamma(I)\cap A_+$.
Furthermore, from (\ref{vert1}) we see that $W^{\lambda_A}(R+it)$ is strictly increasing on $(-\infty,t_1]$
and is strictly decreasing on $[t_2,\infty)$.  By Lemma~\ref{lampotconstant}, we can take 
$I=[c,d]$ and so $ \support\lambda_A =  A_+$. 
\end{proof}

Even in the case when $A$ is a circle in $H^+$ (so that $\Gamma(A)$ is a torus in $\R^3$), it is difficult to 
directly verify  the  hypothesis (b) of Theorem~\ref{convexlemma1}.   
  We next develop sufficient conditions for (b)  that, at least in the case $A$ is a circle,
  are relatively simple to verify. 
  
   For $w\in H^+$ and $t\in[a,b]$, let $r_w(t):=|\gamma(t)-w|$, and $s_w(t):=r_w(t)+r_{w_*}(t)$. 
Assuming $\gamma$ is twice differentiable at $t$ we have
\begin{equation}
\frac{d^2}{dt^2} K(\gamma(t),w)=\frac{-s_w''(t)s_w(t)+s_w'(t)^2}{s_w(t)^2} \qquad (t\in [a,b]).
\end{equation}
Then for fixed $w$, we have that $K(\gamma(t), w)$ is strictly convex  on any interval where $s_w''<0$.   Let $u_w(t)$ denote the unit vector $(\gamma(t)-w)/r_w(t)$.  
Differentiating the dot product $r_w(t)^2 = (\gamma(t)-w)\cdot(\gamma(t)-w)$ we obtain
\begin{eqnarray}
r_w'(t)&=& \gamma'(t)\cdot u_w(t), \nonumber \\
u_w'(t)&=& \left(\gamma'(t)-(\gamma'(t)\cdot u_w(t))u_w(t)\right)/r_w(t), \text{ and} \label{circ1}\\
r_w''(t)&=&\gamma''(t)\cdot u_w(t)+\left(|\gamma'(t)|^2-(\gamma'(t)\cdot u_w(t))^2\right)/r_w(t).\label{circ2}
\end{eqnarray} 
In the event that  $\gamma$ is   parametrized by arclength the above equations can be simplified.   In this case    $|\gamma'(t)|=1$. We further assume that $\gamma''(t)\neq 0$ for any $t\in [a,b]$.   Then $T(t)=\gamma'(t)$ denotes the unit tangent vector, $\kappa(t)=|T'(t)|$ denotes the curvature, and 
 $N(t)=T'(t)/|T'(t)|=\gamma''(t)/\kappa(t)$ denotes the unit normal vector to 
the curve  $\gamma$  for $t\in [a,b]$.  
 Substituting these expressions into (\ref{circ1}) and (\ref{circ2}) we obtain
\begin{eqnarray}
 r_w''(t)&=&\gamma''(t)\cdot u_w(t)+\gamma'(t)\cdot u_w'(t) \label{circ3}\\
 &=& (N(t)\cdot u_w(t))\left[\kappa(t)+\frac{N(t)\cdot u_w(t)}{r_w(t)}\right].  \label{circ4}
 \end{eqnarray}
 
 From this last representation deduce the following. 
 
 \begin{lemma}\label{kappa}
Let $\gamma:[a,b]\to H^+$ be a twice differentiable curve  such that $|\gamma'(t)|=1$
and $\gamma''(t)\neq 0$ for all $t\in [a,b]$.  
Suppose that  for all $s,t \in [a,b]$, $s\neq t$,  and $w\in \{\gamma(s), \gamma(s)_*\}$ we have
 \begin{equation}\label{kappa1}
 N(t)\cdot u_w(t)<0 \text{ and } \left[\kappa(t)+\frac{N(t)\cdot u_w(t)}{r_w(t)}\right] >0.
 \end{equation}
 Then   $\gamma$ satisfies hypothesis (b) of  Theorem~\ref{convexlemma1}.  
 \end{lemma}

  We now apply Lemma~\ref{kappa} to the case when $A_+$ is a subset of a circle. 
  
 \begin{corollary} \label{circsupp}  Suppose $C\subset \C$ is a circle of radius $r>0$
 and center $a$ with $\Re[a]>0$ and suppose $A$ is a compact set in $H^+$ such that
 $A_+\subset C_+$. Then     $\support \lambda_A=A_+^\theta:=A_+\cap \{a+re^{it} \mid |t|\le \theta\}$ for some $\theta\in [0,\pi/2]$.   In particular, if $A_+$ is a circular
 arc contained in $C_+$, then so is $\support \lambda_A$;
 consequently, $\support  \mu_{\Gamma(A)}$ is connected. 
\end{corollary}
{\noindent \bf Remark:}
In the case when $\Gamma(A)$ is a torus (that is, if $A=C$), it follows from Corollary~\ref{circsupp} that 
$\support \mu_{\Gamma(A)}$ is a connected strip of $\Gamma(A)$ of the form $\Gamma(C_+^\theta)$ for some $\theta\in [0,\pi/2]$.

\begin{proof}
Without loss of generality we may assume that $C$ has radius $r=1$ and center $a=R$ for some $R>0$.  We then consider the parametrization of $C$ given by $\gamma(t):=a+e^{it}$ for  $ t\in[-\pi/2, \pi/2]$.   By direct calculation (assisted by Mathematica) we find, for $w=\gamma(s)$, 
$$
N(t)\cdot u_w(t)=-\left|\sin   \frac{s-t}{2} \right| \text{ and } \left[\kappa(t)+\frac{N(t)\cdot u_w(t)}{r_w(t)}\right]=\frac{1}{2},$$
and for  $w=\gamma(s)_*$ we find
$$N(t)\cdot u_{w}(t)=-\frac{   2 R \cos t+\cos (s+t)+1}{\sqrt{(2 R+\cos
  s+\cos t)^2+(\sin s-\sin t)^2}}
$$
and 
$$
 \left[\kappa(t)+\frac{N(t)\cdot u_{w}(t)}{r_{w}(t)}\right]=\frac{1}{2}+\frac{2 R (R+\cos s)}{(2 R+\cos s+\cos
   t)^2+(\sin s-\sin t)^2}.
$$
Then it is easy to verify that the inequalities (\ref{kappa1}) hold for both $w=\gamma(s)$ and for $w=\gamma(s)_*$ for all $s, t\in [-\pi/2,\pi/2]$ with $s\neq t$.
\end{proof}

\section{Kernel in limit $R\to \infty$}
One might well conjecture looking at Figure~1 and in light of Theorem~\ref{Aplus} or Corollary \ref{vertprop} that 
for the case of the circle $A=\{z \mid  |z-R|=1\}$, $R>0$, the support of $\lambda_A$ is the right-half circle  $A_+$, or equivalently, that the support of the equilibrium measure on the torus 
$\Gamma(A)$ is the portion of its surface with positive curvature.  
However, as we see in the limiting case  $R\to \infty$, this is not correct. 

Define the kernels
 $K_R:H^+\times H^+\to \R$, $R>0$,  and $K_{\infty}:H^+\times H^+\to \R$
  by 
  \begin{eqnarray}\label{KR}
  K_R(z,w)&:=&2R \left(K(R+z,R+w)+\log R \right),\\
  K_{\infty}(z,w)&:=&-(\Re[z-w_*]+|z-w|). \label{Kinf}
  \end{eqnarray}
Using 
 $$
 \frac{|z-w|+|2R+z-w_*|}{2R}=1+\frac{\Re[z-w_*]+|z-w|}{2R}+\mathcal{O}(R^{-2})
 $$
 we obtain 
\begin{align*}
K_R(z,w)&= -2R\log\frac{|z-w|+|2R+z-w_*|}{2R}\\
&=-2R\log\left(1+\frac{\Re[z-w_*]+|z-w|}{2R}+\mathcal{O}(R^{-2})\right)\\
&=-(\Re[z-w_*]+|z-w|)+\mathcal{O}(R^{-1})
\end{align*}
and hence
$$\lim_{R\to\infty}K_R(z,w)=K_\infty(z,w),$$  where the convergence is uniform on 
compact subsets of  $H^+\times H^+$.   We let $J_{K_R}(\mu)$ and $J_{K_\infty}(\mu)$
denote the associated energy integrals defined for compactly supported measures $\mu\in \mathcal{M}(H^+)$.  

 From the definition of $K_R$ we see that
the equilibrium measure $\lambda_A^R$ on a compact set $A\subset H^+$ with respect to the kernel $K_R$ is
equal to $\lambda_{A+R}(\cdot+R)$, that is,  $\lambda_A^R(B)=\lambda_{A+R}(B+R)$ where,
for a set $B\subset H^+$ and $R>0$,   $B+R$ denotes the translate $\{b+R \mid  b\in B\}$.

\subsection{The existence and uniqueness of an equilibrium measure for $K_\infty$.}

The   weak-star compactness of $\mathcal{M}(A)$ and the continuity of $J_{K_\infty}$ imply the existence of a measure $\lambda^\infty_A\in\mathcal{M}(A)$  such that
$J_{K_\infty}(\lambda^\infty_A)=\inf_{\mu\in\mathcal{M}(A)}J_{K_\infty}(\mu)$.

We follow arguments developed in \cite{Bjorck}  to prove the uniqueness of $\lambda^\infty_A$. 
First, note that $K_\infty(z,w)=-k_1(z,w)-k_2(z,w)$ where $k_1(z,w):=|z-w|$ and $k_2(z,w)=\Re[z]+\Re[w]$
and so 
$$J_{K_\infty}(\mu)=-I^*_1(\mu)-I^*_2(\mu),  $$  
where $I^*_1$ and $I^*_2$ are the energy integrals associated with the kernels
$k_1$ and $k_2$, respectively.     We need the following lemma of Frostman (\cite{Frostman}, also see \cite[Lemma 1]{Bjorck}).
\begin{lemma}\label{Frost}
Suppose $\nu$ is a compactly supported signed Borel measure on $H^+$ such that
$\int d\nu=0$ and $I^*_1(\nu)\ge 0$.  Then $\nu\equiv 0$.
\end{lemma}

For compactly supported Borel measures $\mu$ and $\nu$ on $H^+$, let
$$J_{K_\infty}(\mu, \nu):= \iint K_\infty(z,w)\, d\mu(z)\, d\nu(w).$$

\begin{lemma} \label{munu} Suppose $A$ is a compact set in $H^+$ and 
$\mu^*\in \mathcal{M}(A)$ satisfies $J_{K_\infty}(\mu^*)=\inf_{\mu\in\mathcal{M}(A)}J_{K_\infty}(\mu)$. 
For any  signed Borel measure $\nu$   with support contained in $A$ such that  $\nu(A)=\int_A d\nu=0$ and  
$\mu^*+\nu\ge 0$, we have $J_{K_\infty}(\mu^*,\nu)\ge 0$.  
\end{lemma}
\begin{proof}
With $\nu$ and $\mu^*$ as above, we have  $\mu^*+\epsilon\  \nu\in  \mathcal{M}(A)$ for $0\le     \epsilon\le 1$ and so 
\begin{equation}\label{JKeps}
J_{K_\infty}(\mu^*)\le J_{K_\infty}(\mu^*+\epsilon\ \nu)=J_{K_\infty}(\mu^*)+2\epsilon J_{K_\infty}(\mu^*,\nu)
+\epsilon^2 J_{K_\infty}(\nu).
\end{equation}
Since (\ref{JKeps}) holds for all $0\le \epsilon\le 1$, then
$J_{K_\infty}(\mu^*,\nu)\ge 0$.
\end{proof}


\begin{theorem}
Suppose $A$ is a compact set in the interior of $H^+$. 
There is a unique equilibrium measure $\lambda^\infty_A$  minimizing 
$J_{K_\infty}(\mu)$ over all $\mu\in\mathcal{M}(A)$.  The support of $\lambda^\infty_A$
is contained in $A_+$.   Furthermore, $\lambda_A^R$ converges  weak-star to $\lambda_A^\infty$
as $R\to \infty$.
\end{theorem}
\noindent
{\bf Remark:}
Recall that $\lambda_A^R$ converges {\em weak-star} to $\lambda_A^\infty$ (and we write
$\lambda_A^R {\ \ \ast \over}\!\!\!\!\to \lambda_A^\infty$)
as $R\to \infty$ means 
that
$$
\lim_{R\to\infty}{\int_{A}{f\,d\lambda_A^R}}=\int_{A}{f\,d\lambda_A^\infty}
$$
for any function $f$ continuous on $A$.

\begin{proof}
Suppose  $\mu^*$ and $\tilde \mu^*$ are measures in $\mathcal{M}(A)$ such that
$J_{K_\infty}(\mu^*)=J_{K_\infty}(\tilde \mu^*)=\inf_{\mu\in\mathcal{M}(A)}J_{K_\infty}(\mu)$.  Then
$\nu:=\tilde \mu^*-\mu^*$ satisfies the hypotheses of Lemma~\ref{munu} and thus 
$J_{K_\infty}(\mu^*,\nu)\ge 0$.  On the  other hand,
$$
J_{K_\infty}( \tilde  \mu^*)=J_{K_\infty}( \mu^*+\nu)=J_{K_\infty}( \mu^*)+2J_{K_\infty}(\mu^*,\nu)
+ J_{K_\infty}(\nu),
$$
which, since $J_{K_\infty}(\mu^*)=J_{K_\infty}(\tilde \mu^*)$, implies that $J_{K_\infty}(\nu)=-2J_{K_\infty}(\mu^*,\nu)\le 0$.  Now, $J_{K_\infty}(\nu)=-I^*_1(\nu)-I^*_2(\nu)=-I^*_1(\nu)$ since
$$I^*_2(\nu)=\iint(\Re[z]+\Re[w])\, d\nu(z)\, d\nu(w)=0.$$  Hence,  $I^*_1(\nu)=-J_{K_\infty}(\nu)=
2J_{K_\infty}(\mu^*,\nu)\ge 0$ and so, by
Lemma~\ref{Frost}, it follows that $\nu\equiv 0$ and thus $\mu^*=\tilde \mu^*$.  

The fact that $\support \lambda^\infty_A\subset A_+$ follows from the observation that $K_{\infty}(z,w)$
is strictly decreasing for $z$ varying along all horizontal rays $[iy,\infty+iy)$ for $y\neq \Im[w]$
and along the  ray $[w,\infty+iv)$ for $v=\Im[w]$, and is constant along the line segment $[iv,w]$.

The weak-star convergence of $\lambda_A^R$ to $\lambda_A^\infty$ follows from the weak-star
compactness of $\mathcal{M}(A)$ and the uniqueness of the equilibrium measure $\lambda_A^\infty$.
\end{proof}

\noindent
{\bf Remarks:} 
\begin{enumerate}
\item The level sets of $K_\infty(\cdot,w)$ are parabolas with focus $w$ and directrix $x=a$ for
$a>\Re[w]$ (in the case $a=\Re[w]$, the level set is the line segment $[iv,w]$ where $v=\Im[w]$).
Notice that these parabolas can also be viewed as arising from the elliptical level curves illustrated in 
Figure~3 by letting the real part of the focus $w_*$ tend to $-\infty$. 
\item One may also consider $K_\infty(z,w)$ on $\C\times \C$ rather than
$H^+\times H^+$ (in effect,  the line $\Re[z]=-\infty$ may be considered the axis of rotation).
\end{enumerate}
Let $W_\infty^\mu$ denote the potential for a measure $\mu\in \mathcal{M}(H^+)$
and  kernel $K_\infty$:
$$W_\infty^\mu(z)=\int_A K_\infty(z,w)\, d\mu(w)\qquad (z\in H^+).$$  
Then  $W_\infty^\mu$ is continuous on $H^+$.  Furthermore, if
$W_\infty^\mu(z)$ is not constant for $z\in \support \mu$, then one may construct a  signed Borel measure $\nu$   with support contained in $A$ such that  $\nu(A)=\int_A d\nu=0$,  
$\mu +\nu\ge 0$, and such that  $J_{K_\infty}(\mu,\nu)< 0$ (cf. \cite{Bjorck}).  
Lemma~\ref{munu} then implies that $J_{K_\infty}(\mu,\nu)$ cannot be minimal,
which gives the following result. 
\begin{lemma}\label{Winf}
The equilibrium potential  $W_\infty^{\lambda_A^\infty}$ satisfies \begin{equation}
\label{Winf1}
W_\infty^{\lambda_A^\infty}(z) \ge J_{K_\infty}(\lambda_A^\infty)\qquad (z\in A)
\end{equation}
with equality if $z\in \support \lambda_A^\infty$.
\end{lemma}

\subsection{Properties of the equilibrium measure for a circle.}

We next consider the support of the $K_\infty$-equilibrium measure in the
 case that $A_+$ is contained in the right-half of a circular arc (as in Corollary~\ref{circsupp}).
  Recall that if $C$ is the circle with center $a$ and radius $r$ and $B\subset C$, we define
  $B^\theta:=B\cap\{a+re^{it} \mid   -\theta\le t\le \theta\}.$ 
\begin{theorem}\label{piover3}
 Suppose $C\subset \C$ is a circle of radius $r>0$
 and center $a$ with $\Re[a]>0$ and suppose $A $ is a non-empty compact set in $H^+$ such that
 $A_+\subset C_+$.  Then     $\support \lambda_A^\infty=A_+^\theta$ for some $\theta\in [0,\pi/2]$.

Furthermore, if $A_+$ is also symmetric about the line $y=\Im[a]$  and $A_+^{  \thetac  }$  is non-empty, then
 $\support \lambda_A^\infty=A_+^\theta$ for some $\theta\in [0,  \thetac  ]$. 
 Moreover,  if $A_+$ is also symmetric about the line $y=\Im[a]$  and $A_+^{  \thetac  }$  is empty, then  $\lambda_A^\infty=(\delta_{a+\zeta}+\delta_{a+\overline{\zeta}})/2$
  where  $\zeta:=re^{i\theta_m}$ and 
 $\theta_m:=\min \{ \theta\ge 0 \mid a+re^{i\theta}\in A_+\}$. 
\end{theorem}

\begin{proof}
Without loss of generality we may assume that $C$ has radius $r=1$ and center $a=0$.  We then consider the parametrization of $C$ given by $\gamma(t):=e^{it}$ for  $-\pi/2\le t\le \pi/2$.  
Then, using $|e^{it}-e^{is}|=2\left|\sin((s-t)/2)\right|$, we find
$$
K_\infty(\gamma(t),\gamma(s))=-\cos(t)-\cos(s)-2\left|\sin  \frac{s-t}{2}\right|\qquad (s,t\in[-\pi/2,\pi/2]).
$$
Differentiating twice with respect to $s$ we obtain
$$\frac{\partial^2}{\partial t^2}K_\infty(\gamma(t),\gamma(s))=\frac{1}{2} \left|\sin \frac{s-t}{2}\right|+\cos (t)$$
which is positive for   $-\pi/2< s,t<\pi/2$.  Then (as in the proof of Corollary~\ref{circsupp}) it follows
that $\support \lambda_A=A_+^\theta$ for some $\theta\in [0,\pi/2]$. 

Now suppose $A_+$ is symmetric about the $x$-axis.  
Then the uniqueness of $\lambda_A^\infty$ shows that
$ \lambda_A^\infty $ is also symmetric about the $x$-axis, that is, 
$ d\lambda_A^\infty (w)=d\lambda_A^\infty (\overline{w})$ for $w\in H^+$.  Thus we have
\begin{eqnarray*}
W_\infty^{\lambda_A^\infty}(\gamma(t)) 
&=&\int_{A_+}K_\infty^s(z,w)\,d\lambda_A^\infty(w),
\end{eqnarray*}
where
$$
K_\infty^s(z,w):=\left(K_\infty(z,w)+K_\infty(z,\overline{w})\right)/2 \qquad (z,w\in H^+).
$$
Then we have 
\begin{align*}
K_\infty^s(&\gamma(t),\gamma(s))\\
&=\begin{cases}
-\cos (s)-\cos (t)-2 \cos \left(s/2\right) \sin \left(t/2\right),& 0\le s<t\le \pi/2,\\
-\cos (s)-\cos (t)-2 \cos \left(t/2\right) \sin \left(s/2\right),& 0\le t<s \le \pi/2.
\end{cases}
\end{align*}
and differentiating with respect to $t$ we obtain
\begin{align}\label{dKdt}\frac{\partial}{\partial t}&K_\infty^s(\gamma(t),\gamma(s))\\
&=\begin{cases}
\sin (t)-\cos \left(s/2\right) \cos \left(t/2\right),& 0\le s<t\le \pi/2,\\
\sin (t)+\sin \left(s/2\right) \sin \left(t/2\right),& 0\le t<s\le \pi/2.
\end{cases}\nonumber
\end{align}
We claim that 
 \begin{equation}\label{dKdtpos}
 \frac{\partial}{\partial t}K_\infty^s(\gamma(t),\gamma(s))>0 \qquad (-\pi/2\le s\le\pi/2,\, t>  \thetac  ).
 \end{equation}
Clearly (\ref{dKdtpos}) holds in the second case of (\ref{dKdt}) when $0< t<s\le \pi/2$.
If $   \thetac  < t \le \pi/2$ and $0\le s<t$, then using the  first case of (\ref{dKdt}), 
$$
\sin (t)-\cos \left(s/2\right) \cos \left(t/2\right)=\cos \left(t/2\right)\left(2\sin \left(t/2\right)-\cos \left(s/2\right)\right)
$$
and   $2\sin \left(t/2\right)-\cos \left(s/2\right) \ge 2\sin \left(t/2\right)-1>0$ for this range
of $s$ and $t$, we see that (\ref{dKdtpos}) holds in this case as well. 
 Hence, we have 
\begin{eqnarray*}
\frac{d}{dt}W_\infty^{\lambda_A^\infty}(\gamma(t))=\int_{A_+}\frac{\partial}{\partial t}K_\infty^s(\gamma(t),w)\,d\lambda_A^\infty(w)>0 \qquad (t>  \thetac  ).
\end{eqnarray*}
Thus,  in light of Lemma~\ref{Winf}, we have $\support \lambda_A^\infty\subset A_+^{  \thetac  }$ if $A_+^{  \thetac  }\neq \emptyset$, while if $A_+^{  \thetac  }= \emptyset$, then $\lambda_A^\infty=(\delta_{a+\zeta}+\delta_{a+\overline{\zeta}})/2$.
\end{proof}


 \subsection{The vertical line segment.}
   In this section we consider sets $A\subset H^+$ such that $A_+$ is contained
   in a vertical line segment $[a+ic,a+id]$  and further suppose
  the endpoints $a+ic$ and $a+id$ are in $A_+$.
   Then   $$K_\infty(a+it,a+is)=-2a -|t-s|\qquad (s,t\in[c,d])$$
which falls into the class of kernels studied in \cite{Bjorck} and it follows from
results there that $\lambda_A^\infty=\left(\delta_{a+ic}+\delta_{a+id}\right)/2$ where 
$\delta_w$ denotes the unit point mass at $w$. 
In particular, for the  ``infinite washer'' in $\R^3$
obtained by rotating $[a+ic,a+id]$ about the $y$-axis and letting
$a\to\infty$, the support of the equilibrium measure degenerates to
two circles.   We contrast this with the finite $R$ case
where, by Corollary~{\ref{vertprop}},  we have $\support \lambda_A^R =A_+$ .

\section{Discrete Minimum Energy Problems on $A\subset H^+$}
Suppose $A\subset H^+$ is compact,   $k:A\times A \to \R_+$ is continuous and nonnegative, and that there is a unique
equilibrium measure $\lambda_{k,A}$ minimizing the $k$-energy 
$$
I_k(\mu):= \iint_{A\times A}  k  (x,y) \ d\mu (x)\ \! d\mu (y) 
$$
over measures $\mu\in\mathcal{M}(A)$.  In this case we say that $k$ is a {\em continuous admissible kernel on $A$}.   In particular, we have in mind the  reduced kernel 
$ K$ as defined in (\ref{Kdef}) or the limiting kernel $K_{\infty}$ as defined in
(\ref{Kinf}).

We consider the following discrete minimum $k$-energy problem.  The arguments in this section closely 
follow those in  \cite[ pp.  160--162]{L}; however, the continuity of $k$ here allows for some simplification. 
For a collection of $N \ge 2$ distinct points
$\omega_N:=\{x_1,\ldots, x_N\}\subset A$, let 
$$
E _k(\omega_N):= 	\sum_{1\leq i\neq j\leq N}{k(x_i,x_j)} =\sum_{i=1}^N\sum_{ {j=1} \atop { j\neq i}}^Nk(x_i,x_j),
 $$
and 
\begin{equation} \label{Ek}\mathcal E_k(A,N):=\inf \{E_k(\omega_N) \mid \omega_N\subset A ,
 |\omega_N |=N\}.
\end{equation} 

Since 
\begin{equation}\label{dminen1}
\mathcal E_k(A,N)\le 	\sum_{1\leq i\neq j\leq N}{k(x_i,x_j)} 
\end{equation} for any configuration of $N$ points $\{x_1,\ldots, x_N\}\subset A$, integrating (\ref{dminen1})
with respect to $d\lambda_{k,A}(x_1)d\lambda_{k,A}(x_2)\cdots d\lambda_{k,A}(x_N)$ we find
$\mathcal E_k(A,N)\le 	N(N-1)I_k(\lambda_{k,A})$ and so  we have
\begin{equation}\label{dminen2}
\frac{\mathcal E_k(A,N)}{N(N-1)}\le 	I_k(\lambda_{k,A}) \qquad (N\ge 2).
\end{equation}

On the other hand, the compactness of $A$ and continuity of $k$ imply that  for each $N\ge 2$ there exists  some {\em  optimal $k$-energy
configuration} $\omega^*_N\subset A$ such that $E_k(\omega^*_N)=\mathcal{E}_k(A,N)$.  
Let $\lambda_{A,N}=\frac{1}{N}\sum_{x\in \omega^*_N}\delta_x\in \mathcal{M}(A)$ (where $\delta_x$ denotes
the unit point mass at $x$).  Then 
\begin{equation}\label{dminen3}
I_k(\lambda_{k,A})\le I_k(\lambda_{A,N})
= \frac{\mathcal{E}_k(A,N)+ \sum_{i=1}^Nk(x_i,x_i)}{N^2}
 \qquad  (N\ge 2).
\end{equation}
Combining (\ref{dminen2}) and (\ref{dminen3}) we have
 \begin{equation}\label{dminen3.5}
\frac{\mathcal{E}_k(A,N)}{N(N-1)}\le I_k(\lambda_{k,A})\le I_k(\lambda_{A,N})
\le \frac{\mathcal{E}_k(A,N)}{N^2} +\frac{\|k\|_{A}}{N}  \quad (N\ge 2),
\end{equation}
where  $\|k\|_{A}:=\sup_{z\in A}k(z,z)$.  Since ${\mathcal{E}_k(A,N)}/{N^2}\le I_k(\lambda_{k,A})<\infty$,  the inequalities in (\ref{dminen3.5}) show that there is some constant $C$ such that $0\le I_k(\lambda_{A,N})- I_k(\lambda_{k,A})\le C/N$ for $N\ge 2$, and so
\begin{equation}\label{dminen4}
I_k(\lambda_{A,N})\to  I_k(\lambda_{k,A})\text{ as }N\to\infty .  
\end{equation}
If $\mu^*$ is a weak-star limit point of the sequence $\{\lambda_{A,N}\}$, then (\ref{dminen4}) shows that
$I_k(\mu^*)=I_k(\lambda_{k,A})$ and so $\mu^*=\lambda_{k,A}$.    By the weak-star compactness of $\mathcal{M}(A)$, any 
subsequence of $\{\lambda_{A,N}\}$ must contain a weak-star convergent subsequence. Hence, we have the following result.
\begin{proposition} \label{DMEthm}
Suppose $A$ is a compact set in $H^+$ and that $k:A\times A \to \R_+$  is a continuous  admissible kernel on $A$. 
For $N\ge 2$, let  $\omega^*_N$ be an optimal $k$-energy
configuration of $N$ points $\{x_1,x_2,\ldots, x_N\}\subset A$.  Then
$
\frac{1}{N}\sum_{i=1}^N\delta_{x_i} {\ \ \ast \over}\!\!\!\!\to \lambda_{k,A}$ as $N\to \infty$.
\end{proposition}

\begin{figure}[tbp]
\begin{center}
(a) \includegraphics[scale=.8]{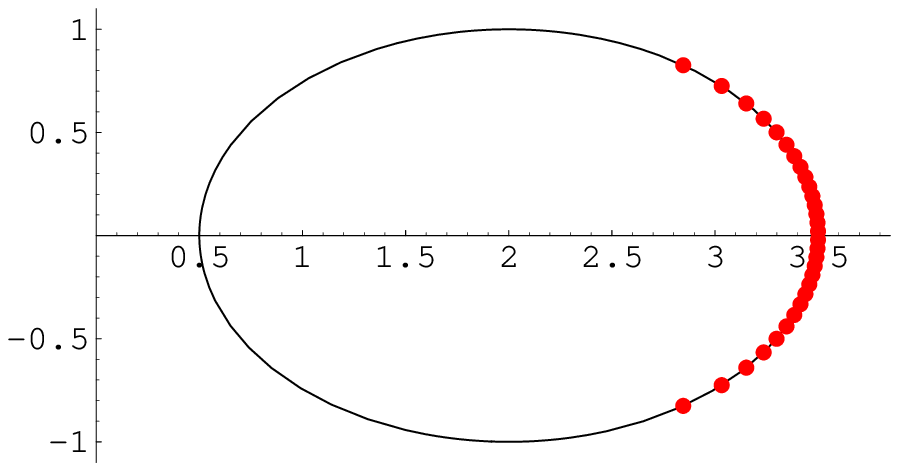}

(b) \includegraphics[scale=.8]{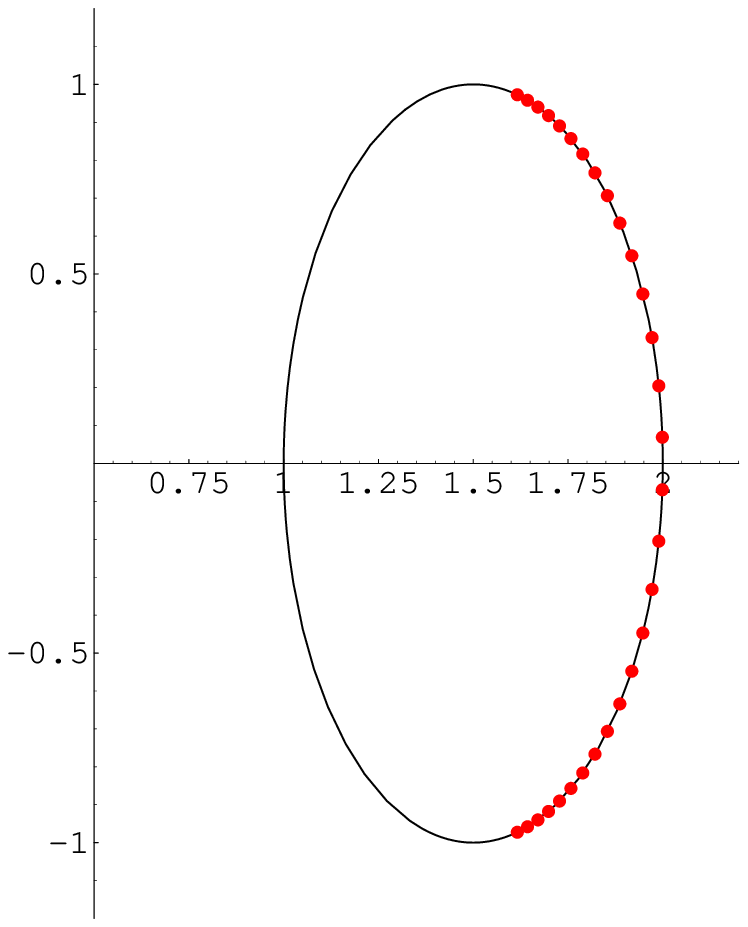}

(c) \includegraphics[scale=.8]{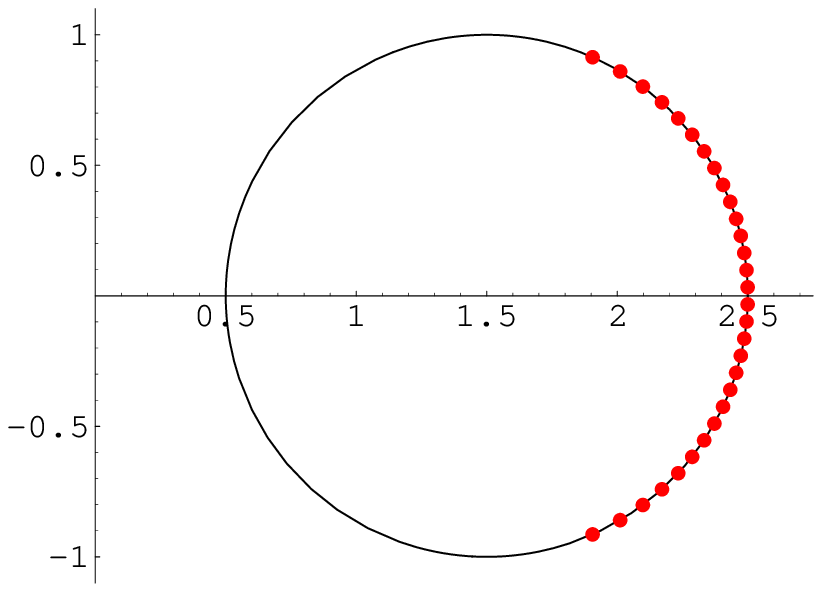}
\caption{Near optimal $K$-energy configurations ($N=30$ points) on various ellipses in $H^+$.}
\label{discfig}
\end{center}
\end{figure}

Figure~\ref{discfig} shows (near) optimal $K$-energy configurations for $N=30$ points restricted to various ellipses in $H^+$.

\medskip
\noindent
{\bf Acknowledgement.}  We thank Rob Womersley for performing the computations and providing the resulting images shown in Figures~1 and 2.  We also extend our appreciation to Johann Brauchart his careful reading of the original manuscript. 



%
%


\newpage

\begin{thebibliography}{99}

\bibitem{BF} J. Benedetto  and M. Fickus,   
{ Finite normalized tight frames}, {\it  Adv. Comput. Math.} {\bf 18} (2003),  357--385.

\bibitem{Bjorck}  G. Bj\"{o}rck, Distributions of positive mass which maximize a certain generalized energy integral, {\it Ark. Mat.} {\bf 3} (1956), 255--269. 



 \bibitem{BorHarSaf05} S. Borodachov, D. P. Hardin, and E. B. Saff, 
On asymptotics of the weighted Riesz energy for rectifiable sets,
submitted (2005).

\bibitem{BNT04} M. Bowick,  D. R. Nelson,  and A. Travesset,
Curvature-induced defect unbinding in toroidal geometries, {\it Phys. Rev. E} {\bf 69},
(2004), 041102--041113.

\bibitem {ConSlo99}
J.H. Conway and N.J.A. Sloane, {\it Sphere Packings, Lattices and Groups,}
Springer Verlag, New York: 3rd ed., 1999.


\bibitem{Frostman} Frostman, O., Potentiel d'\'{e}quilbre et capacit\'{e} des ensembles, {\it Medd. Lunds Univ.  Mat. Sem.} {\bf 3} (1935).


  
 \bibitem{HarSaf04} D.P. Hardin and E.B. Saff, Minimal Riesz energy point configurations
for rectifiable $d$-dimensional manifolds, {\it Adv.  Math.} {\bf 193} (2005), 174--204.

 \bibitem {HarSaf04N}
D.P. Hardin and E.B. Saff, Discretizing manifolds via minimum energy
points, {\it Notices of the AMS.}  {\bf 51} (2004), no. 10, 1186--1194.

\bibitem {KuiSaf98}
A.B.J. Kuijlaars and E.B. Saff, Asymptotics for minimal discrete energy on the sphere, {\it Trans. Amer. Math. Soc.}  {\bf 350} (1998), no. 2, 523--538.

\bibitem {MatGSMES}
P. Mattila, {\it Geometry of sets and measures in Euclidian spaces. Fractals and Rectifiability,} Cambridge Univ. Press, 1995, 344 pages.
\bibitem {L}
N.S. Landkof, {\it Foundations of modern potential theory.}
Springer-Verlag, Berlin-Heidelberg-New York, 1972, 426 pages.

\bibitem{ST}
E.B. Saff and V. Totik, {\it Logarithmic potentials with external fields,}
Springer-Verlag, Berlin-Heidelberg-New York, 1997, 508 pages.

\bibitem{SloWom} I.H. Sloan and R.S. Womersley, Extremal systems of points and numerical integration on the Sphere, {\it Adv. Comp. Math.} {\bf 21} (2004), 102--125. 

\bibitem{S} {S. Smale},
{ Mathematical problems for the next century},
{\it Mathematical Intelligencer}, {\bf 20} (1998), 7--15.

\vspace*{.5in}

\end{thebibliography}
 \end{document}